\documentclass[aps,twocolumn]{revtex4}
\newcommand{\mb}[1]{ { \mbox{\boldmath{$#1$}}}  } 
\usepackage{epsf} 
\begin{document} 
\title{Meservey-Tedrow-Fulde effect in a quantum dot embedded \\
       between metallic and superconducting electrodes}
  
\author{T.\ Doma\'nski$^{1}$, A.\ Donabidowicz$^{1}$,  
        and K.I.\ Wysoki\'nski$^{1,2}$} 
 
\affiliation{$^{1}$Institute of Physics,  
             M.\ Curie Sk\l odowska University,  
             20-031 Lublin, Poland\\ 
	     $^{2}$Max Planck Institut f\"ur 
             Physik komplexer Systeme,  
             D-01187 Dresden, Germany} 
 
\begin{abstract} 
Magnetic field applied to the quantum dot coupled between one 
metallic and one superconducting electrode can produce a similar 
effect as has been experimentally observed by Meservey, Tedrow 
and Fulde [Phys.\ Rev.\ Lett.\ {\bf 25}, 1270 (1970)] for the 
planar normal metal -- superconductor junctions. We investigate  
the tunneling current and show that indeed the square root 
singularities of differential conductance exhibit the Zeeman 
splitting near the gap edge features $V = \pm\Delta/e$. Since 
magnetic field affects also the in-gap states of quantum dot 
it furthermore imposes a hyperfine structure on the anomalous 
(subgap) Andreev current which has a crucial importance for 
a signature of the Kondo resonance.
\end{abstract} 
\date{\today}

\maketitle 
 
\section{Introduction}
 
Already in early days of the tunneling spectroscopy it has been 
shown that magnetic field $B$ (which couples to spin of the charge 
carriers) is in superconductors responsible for splitting the square 
root singularities of the tunneling conductance \cite{Meservey-70} by 
the Zeeman energy $2\mu_{B}B$, where $\mu_B$ is the Bohr magneton. 
This Meservey-Tedrow-Fulde (MTF) effect has been observed 
experimentally in the thin superconducting aluminum films applying 
parallel magnetic field so that orbital diamagnetic effects could 
be avoided. Similar qualitative results have been recently noticed 
in the measurements of $c$-axis tunneling for the layered high 
temperature superconducting compounds \cite{Alvarez06}. 

We argue that the MTF effect should be also feasible in various 
nanostructures consisting of a quantum dot (QD) placed between 
one metallic and one superconducting electrode. Zero-dimensional 
character of QDs in a natural way eliminates the influence of 
orbital effects therefore magnetic field would affect the charge 
transport only through the Zeeman term. This can in turn 
manifest itself in the differential conductance. Roughly speaking, 
the charge current flows if an external bias $V$ exceeds the energy 
gap $\Delta$ (necessary to break the Cooper pairs into individual 
electrons) thereof the resulting conductance has a low voltage 
onset near the gap edges $eV=\pm \Delta$. In presence of a magnetic 
field these gap edge singularities are going to split (see section III). 

More detailed analysis of the charge tunneling \cite{Blonder} 
involves however also the additional (anomalous) channels due 
to mixing of the particle and hole excitations in superconductors. 
In particular, even at subgap voltages $|eV|\leq \Delta$ the 
mechanism of Andreev reflections provides a finite contribution 
to the conductance. Since the Andreev mechanism is very sensitive 
to location of the in-gap QD states \cite{Beenakker92,Claughton95,
Yeyati97,Zhao99,Sun99} and the on-dot correlations \cite{Fazio-98,
Clerk-00,Cuevas-01,Krawiec-04,Oguri-04,Tanaka-07,Domanski-07,
Taddei-07,Gezzi08,Hecht-08}  we shall explore the influence 
of magnetic field on such subgap conductance. In section IV 
we discuss a hyperfine structure for the Andreev conductance 
neglecting the correlations. In the next section V we extend 
our study taking into account a finite value of the on-dot 
repulsion $U$. We show that appearance of the low temperature
Kondo resonance enhances the zero bias conductance and this
feature undergoes the Zeeman splitting when magnetic field 
is applied.

As concerns some practical aspects, there have been considered the 
proposals for using the magnetic field tuned Andreev scattering as 
an efficient cooling mechanism in two dimensional electron gas 
- superconductor nanostructure \cite{Giazotto06}. There is also 
considered a possibility to use the, so called, Andreev quantum 
dot as a magnetic flux detector \cite{Sadovsky07}.

\section{The model}

For a general description of transport phenomena through a nanoscopic
island placed between external leads one should consider a quantized 
multilevel structure of QD \cite{Aleiner-02}. However, in the case 
when a level spacing is smaller in comparison to QD hybridization 
with the electrodes one can restrict to a simplified picture 
of the Anderson model \cite{Claughton95,
Yeyati97,Fazio-98,Clerk-00,Cuevas-01} 
\begin{eqnarray} 
\hat{H} &=& \hat{H}_{N} + \hat{H}_{S} + \sum_{\sigma}  
\epsilon_{d,\sigma} \hat{d}^{\dagger}_{\sigma} \hat{d}_{\sigma}  
+  U \; \hat{n}_{d \uparrow} \hat{n}_{d \downarrow}  
\nonumber \\
&+& \sum_{{\bf k},\sigma } \sum_{{\beta}=N,S}  
\left( V_{{\bf k} \beta} \; \hat{d}_{\sigma}^{\dagger}  
\hat{c}_{{\bf k} \sigma \beta } + V_{{\bf k} \beta}^{*}  
\; \hat{c}_{{\bf k} \sigma, \beta }^{\dagger} \hat{d}_{\sigma} 
\right)  .
\label{model} 
\end{eqnarray} 
Operators $d_{\sigma}$ ($d_{\sigma}^{\dagger}$) denote the annihilation  
(creation) of electron whose energy level is $\varepsilon_{d,\sigma}$ 
and $U$ is the on-dot Coulomb repulsion between opposite spin electrons. 
The last terms describe hybridization of QD with the normal 
($\beta\!=\!N$) and superconducting ($\beta\!=\!S$) electrodes.  
Magnetic field eventually shifts the QD level by $\varepsilon_{d,\sigma}
=\varepsilon_{d} - g_{\sigma}\mu_{B}B$, where the spin-dependent
coefficients are defined as $g_\uparrow\!=\!1$ and $g_\downarrow\!=\!-1$. 
 
Hamiltonian of the normal (metallic) lead is taken as $\hat{H}_{N} 
\!=\! \sum_{{\bf k},\sigma} \xi_{{\bf k}N}^{\sigma} \hat{c}_{{\bf k} 
\sigma N}^{\dagger} \hat{c}_{{\bf k} \sigma N}$  whereas for the 
superconducting electrode we choose the usual BCS form 
$\hat{H}_{S} \!=\!\sum_{{\bf k},\sigma}  \xi_{{\bf k}S}^{\sigma}
\hat{c}_{{\bf k} \sigma S }^{\dagger}  \hat{c}_{{\bf k} \sigma S} 
\!-\! \sum_{\bf k} \left( \Delta  \hat{c}_{{\bf k} \uparrow S }
^{\dagger} \hat{c}_ {-{\bf k} \downarrow S }^{\dagger} + 
\mbox{h.c.} \right)$ with an isotropic energy gap $\Delta$. 
The relative energies $\xi_{{\bf k}\beta}^{\sigma}\!=\!
(\varepsilon_{{\bf k}\beta} \!-\!g _{\sigma} \mu_{B}B)
\!-\!\mu_{\beta}$ are measured from the chemical potentials 
$\mu_{\beta}$. We shall focus on the wide band limit 
$|V_{{\bf k}\beta}| \!  \ll \! D$ (where $-D\!\leq\!
\varepsilon_{{\bf k}\beta} \!  \leq \! D$) and consider a small 
external voltage $V$, which detunes the chemical potentials 
by $\mu_{N}\!-\!\mu_{S}=eV$ inducing the charge flow through 
N-QD-S junction. We assume $|eV|$ to be much smaller than 
level spacings typical for the realistic QDs \cite{Aleiner-02}
so that applicability of the model (\ref{model}) can be justified. 
 
Let us start by establishing the QD Green's function in 
the equilibrium situation, i.e.\ for $V\!=\!0$. Fourier transform 
of the retarded Green's functions can be formally expressed 
by the Dyson equation  
\begin{eqnarray} 
{\mb G}_{\sigma}(\omega)^{-1} &\equiv &
\left[ \begin{array}{cc}  
\langle\langle \hat{d}_{\sigma} ;\hat{d}_{\sigma}^{\dagger}  
\rangle\rangle_{\omega} & \langle\langle\hat{d}_{\sigma}; 
\hat{d}_{\sigma} \rangle\rangle_{\omega} \\ \langle\langle  
\hat{d}_{-\sigma}^{\dagger}; \hat{d}_{\sigma}^{\dagger}  
\rangle\rangle_{\omega} & \langle \langle \hat{d}_{-\sigma} 
^{\dagger} ;\hat{d}_{-\sigma} \rangle\rangle_{\omega}  
\end{array}\right]^{-1} \label{GF}  \\ &=&  
\left[ \begin{array}{cc}  
\omega\!-\!\varepsilon_{d,\sigma} &  0 \\ 0 &  
\omega\!+\!\varepsilon_{d,-\sigma}\end{array}\right]  
- {\mb \Sigma}_{d,\sigma}^{0}(\omega)  
- {\mb \Sigma}_{d,\sigma}^{U}(\omega)  
\nonumber
\end{eqnarray} 
where ${\mb \Sigma}_{d,\sigma}^{0}$ denotes the selfenergy of 
noninteracting QD ($U\!=\!0$) and  ${\mb  \Sigma}_{d,\sigma}^{U}$ 
accounts for the correlation effects. For a simple understanding 
of the MTF effect it would be helpful to focus first on the 
uncorrelated QD when the selfenergy is known exactly. 
Further corrections due to ${\mb \Sigma}_{d,\sigma}^{U}$ 
contribute a renormalization of the spectral function 
\cite{Tanaka-07} whose impact on the charge transport will 
be discussed separately in section V.
 
For convenience we introduce the hybridization coupling 
$\Gamma_{\beta} \equiv  2\pi\sum_{\bf k} |V_{{\bf k}\beta}
|^{2} \delta(\omega\!-\!\varepsilon_{{\bf k\beta}})$ and 
define the following spin-dependent energy $\tilde{\omega}
_{\sigma}\!=\!\omega\!+\!g_{\sigma} \mu_{B}B$. Imaginary 
part of the selfenergy ${\mb \Sigma}_{d,\sigma}^{0}$ for 
$|\tilde{\omega}_{\sigma}|\!\leq\!\Delta$ is given by 
$\mbox{Im}{\mb \Sigma}_{d,\sigma}^{0}(\omega)=-\frac{
\Gamma_{N}}{2}{\mb 1}$ while at large energies $D \! > 
\! |\tilde{\omega}_{\sigma}| \! > \! \Delta$ it takes 
the following form \cite{Krawiec-04,Domanski-07}
\begin{eqnarray} 
&&\mbox{Im}{\mb \Sigma}_{d,\sigma}^{0}(\omega)= \nonumber \\
&&-\frac{1}{2} \left[ \begin{array}{cc}  
\Gamma_{N}+\Gamma_{S}\frac{|\tilde{\omega}_{\sigma}|} 
{\sqrt{\tilde{\omega}_{\sigma}^{2}-\Delta^{2}}} & 
\Gamma_{S} \frac{\Delta \; \mbox{sgn}(\tilde{\omega}_{\sigma})} 
{\sqrt{\tilde{\omega}_{\sigma}^{2}-\Delta^{2}}} \\ 
\Gamma_{S} \frac{\Delta \; \mbox{sgn}(\tilde{\omega}_{\sigma})} 
{\sqrt{\tilde{\omega}_{\sigma}^{2}-\Delta^{2}}} & 
\Gamma_{N}+\Gamma_{S}\frac{|\tilde{\omega}_{\sigma}|} 
{\sqrt{\tilde{\omega}_{\sigma}^{2}-\Delta^{2}}}  
\end{array} \right] . 
\label{imag_free} 
\end{eqnarray} 
The corresponding real parts can be determined using the 
Kramers-Kr\"onig relations. 
 
Imaginary part of the selfenergy ${\mb \Sigma}_{d,\sigma}^{0}$  
has thus the square root singularities at energies $\omega\!=\!
\pm\Delta \pm \mu_{B}B$, so in presence of magnetic field there 
are altogether 4 such points. They show up as kinks in 
the spectral function $\rho_{d}(\omega) =\sum_{\sigma} 
\rho_{d,\sigma}(\omega)$, where 
\begin{eqnarray} 
\rho_{d,\sigma}(\omega) = - \frac{1}{\pi} \mbox{Im}  
\langle\langle \hat{d}_{\sigma} ;\hat{d}_{\sigma}^{\dagger}  
\rangle\rangle_{\omega+i0^{+}} . 
\label{rho_sigma}
\end{eqnarray} 
We shall see below that appearance of such characteristic 
points leads to the MTF effect observed in the tunneling 
conductance.

\section{Meservey-Tedrow-Fulde effect}
 
To compute the tunneling current we adopt the formalism outlined 
in the previous studies \cite{Fazio-98,Sun99,Krawiec-04} extending  
it here on a situation with the spin sensitive transport due to 
magnetic field. The steady charge current is defined as $I(V)\!=
\! -e\frac{d}{dt}\langle \sum_{{\bf k},\sigma} \hat{c}_{{\bf k}  
\sigma N}^{\dagger} \hat{c}_{{\bf k} \sigma N} \rangle\!=\! 
e\frac{d}{dt}\langle \sum_{{\bf k},\sigma} \hat{c}_{{\bf k}  
\sigma S}^{\dagger} \hat{c}_{{\bf k} \sigma S} \rangle$. We 
carry out the time derivative and determine the expectation 
value using the nonequilibrium Keldysh Green's functions. 

\begin{figure} 
{\epsfxsize=9cm \centerline{\epsffile{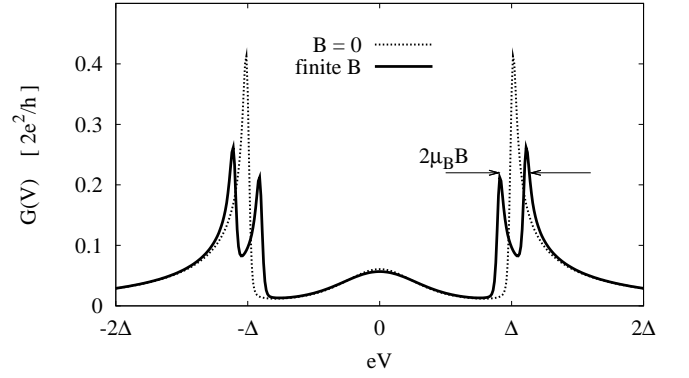}}} 
\caption{The differential conductance $G(V)$ versus  
bias voltage $V$ for N-QD-S junction. Notice a splitting  
of the gap-edge singularities around $eV\!=\!\pm\Delta$  
induced by magnetic field $B$. We used for computations  
$\varepsilon_{d}\!=\!0$, $U\!=\!0$, $\Gamma_{N}\!=\! 
\Delta$, $\Gamma_{S}\!=\!0.1\Delta$, $T=0.01\Delta$  
assuming $\Delta\!=\!0.1D$.} 
\label{MT_plot} 
\end{figure} 

In analogy to the standard Blonder-Tinkham-Klawijk theory 
\cite{Blonder} we express the current as composed 
of two contributions  
\begin{eqnarray} 
I(V) = I_{1}(V) + I_{A}(V) . 
\label{total} 
\end{eqnarray} 
The first part $I_{1}(V)$ stands for a contribution which 
at low temperatures appears practically outside the energy 
gap $|eV|\!\geq\!\Delta$. Its magnitude is expressed by 
the Landauer-type formula 
\begin{eqnarray} 
I_{1}(V) = \frac{e}{h} \sum_{\sigma} \int d\omega \;  
T_{1,\sigma}(\omega) \; \left[ f(\omega\!+\!eV)\!-\! 
f(\omega) \right] , 
\end{eqnarray} 
where $f(\omega)\!=\!\left[1+\mbox{exp}(\omega/k_BT) \right]
^{-1}$. The transmittance $T_{1,\sigma}(\omega)$ is nonvanishing
only outside the energy gap $|\tilde{\omega}_{\sigma}| \geq 
\Delta$ and is given by the following parts of the retarded 
Green's functions \cite{Sun99,Krawiec-04} 
\begin{eqnarray} 
T_{1,\sigma}(\omega) &=& \frac{\Gamma_{N} \Gamma_{S} \;  
|\tilde{\omega}_{\sigma}|}{\sqrt{\tilde{\omega}_{\sigma}^{2} 
-\Delta^{2}}}  \left( \left| \langle \langle \hat{d}_{\sigma};  
\hat{d}_{\sigma}^{\dagger} \rangle \rangle_{\omega}  
\right|^{2} + \left| \langle \langle \hat{d}_{\sigma};  
\hat{d}_{-\sigma} \rangle \rangle_{\omega} \right|^{2}  
\right) \nonumber \\ 
&-& \frac{2 \Gamma_{N} \Gamma_{S} \Delta}{\sqrt{\tilde 
{\omega}_{\sigma}^{2}-\Delta^{2}}}  \; \mbox{Re} \left\{  
\langle \langle \hat{d}_{\sigma}; \hat{d}_{\sigma}^{\dagger}  
\rangle \rangle_{\omega} \;\; \langle \langle \hat{d}_{\sigma};  
\hat{d}_{-\sigma} \rangle \rangle_{\omega}^{*} \right\} . 
\label{T1} 
\end{eqnarray} 
%


The second part in (\ref{total}) originates from the mechanism 
of Andreev reflections \cite{Blonder,Fazio-98,Krawiec-04} 
\begin{eqnarray} 
I_{A}(V) = \frac{e}{h} \sum_{\sigma} \int d\omega  
T_{A,\sigma}(\omega) \left[ f(\omega\!+\!eV)\!-\! 
f(\omega\!-\!eV) \right] . 
\end{eqnarray} 
Its transmittance is finite even inside the energy gap 
\cite{Sun99,Krawiec-04} 
\begin{eqnarray} 
T_{A,\sigma}(\omega) = \Gamma_{N}^{2}  
\left| \langle \langle \hat{d}_{\sigma}; \hat{d}_{-\sigma}  
\rangle \rangle_{\omega} \right|^{2}  .
\label{T_A} 
\end{eqnarray} 
Physically such process occurs when an incident electron from 
$N$ electrode (of arbitrary energy) is converted into a pair 
on QD (with a simultaneous  reflection of a hole) and  
it propagates in $S$ electrode as a Cooper pair. This 
anomalous Andreev current is closely related to the 
off-diagonal order parameter induced in the QD 
(proximity effect) \cite{Domanski-07,Tanaka-07}.  

Figure (\ref{MT_plot}) illustrates the influence of magnetic  
field on the total differential conductance $G(V)\!=\!\frac{d} 
{dV}I(V)$ obtained for N-QD-S junction. We clearly notice 
the Zeeman splitting  of the square root singularities resembling 
the former experimental observation for N-I-S (I-insulator) 
junction \cite{Meservey-70}. However,  in a present case 
the conductance does not saturate to a finite value far 
outside the gap $|eV| \gg \Delta$ because the QD spectrum 
spreads only nearby $\varepsilon_{d}$ (usually in realistic
multilevel QDs there would be seen the quantum oscillations 
of $G(V)$ \cite{Aleiner-02}). The in-gap features related 
to the Andreev current are discussed in the next section. 
 
\section{Magnetic field effect on the Andreev current}
 
\begin{figure} 
{\epsfxsize=7.5cm \centerline{\epsffile{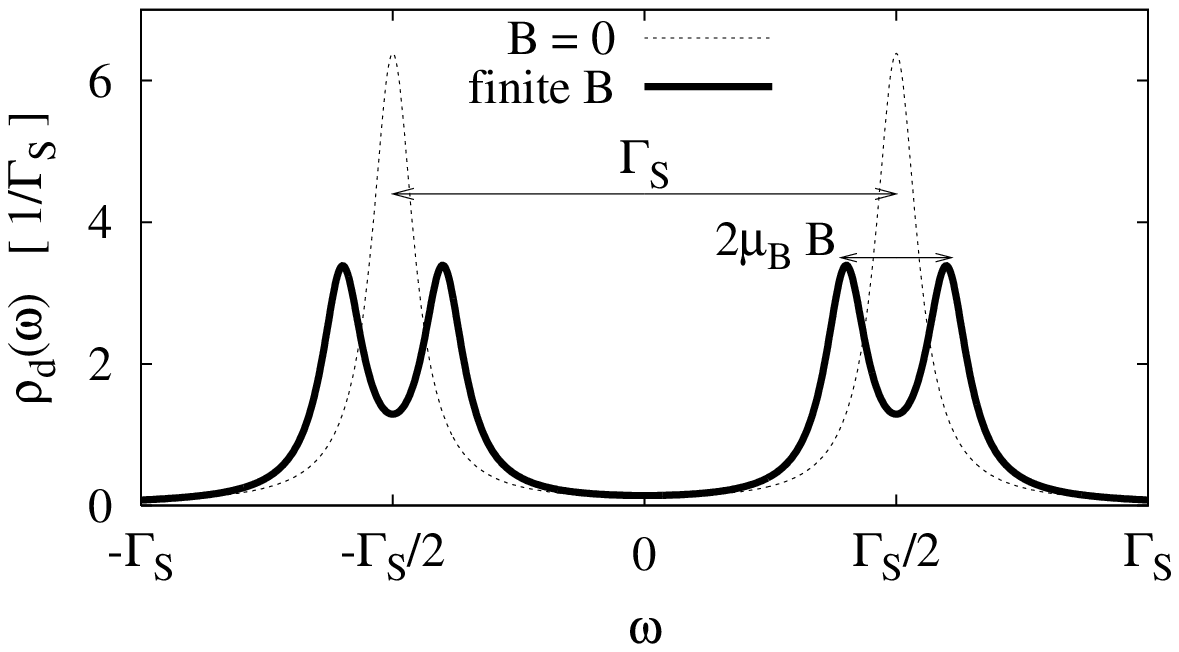}}} 
{\epsfxsize=8cm \centerline{\epsffile{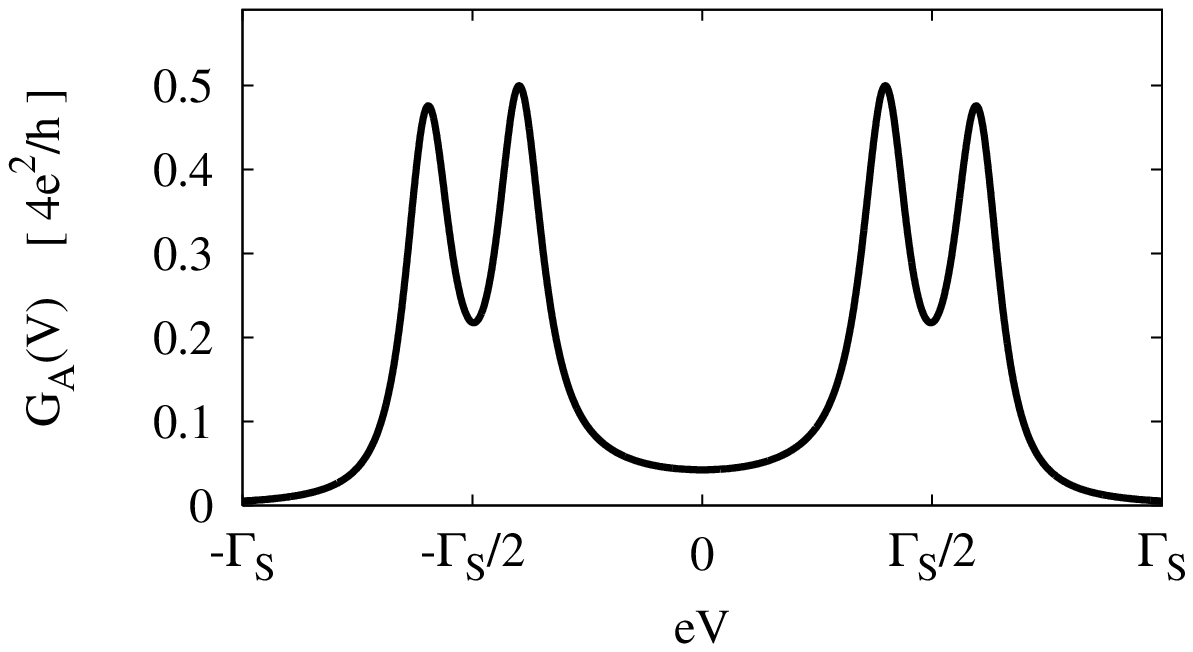}}} 
\caption{Zeeman splitting of the bound Andreev states for  
the QD located in the center of superconducting gap  
$\varepsilon_{d}=0$. Upper panel illustrates the density  
of states $\rho_{d}(\omega)$ and the bottom figure shows  
differential conductance of the in-gap current. For 
computations we used $\Gamma_{N}\!=\!0.1\Gamma_{S}$, 
$\mu_{B}B\!=\!0.1\Gamma_{S}$ assuming $\Gamma_{S}=0.01D$ 
and $U\!=\!0$.} 
\label{bound_states} 
\end{figure} 

\begin{figure}[t]
\epsfxsize=10cm \centerline{\epsffile{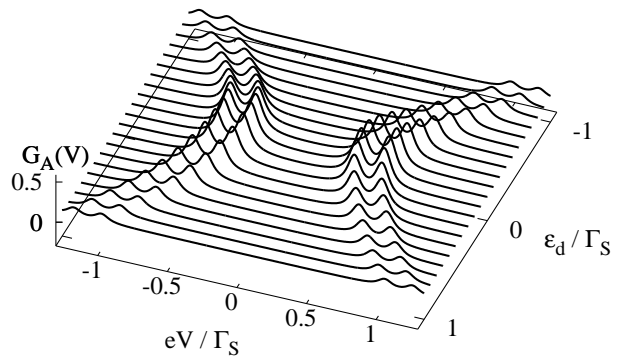}} 
\caption{Differential conductance $G_{A}(V)$ of the in-gap  
Andreev current as a function of the bias voltage $V$ and  
the QD level $\varepsilon_{d}$. We used for computations  
$\Gamma_{S}=0.01D$, $\Gamma_{N}=0.1\Gamma_S$, $T=0.01\Gamma_{S}$  
and set the magnetic field $\frac{1}{2}\mu_{B}B=0.1\Gamma_{S}$. 
The conductance is expressed in units of $4e^{2}/h$.} 
\label{GA_vs_ed} 
\end{figure} 

The mechanism of Andreev reflections transmits the charge current 
even for the subgap voltages. To focus solely on this anomalous  
current it is convenient to consider the extreme limit $\Delta  
\rightarrow \infty$ as proposed by Tanaka et al \cite{Tanaka-07}.  
In such case $I_{1}$ can be completely discarded from our analysis.  
Using (\ref{imag_free}) we obtain the selfenergy 
${\mb \Sigma}_{d,\sigma}^{0}$  simplified to 
\cite{Fazio-98,Domanski-07,Tanaka-07} 
\begin{eqnarray} 
{\mb \Sigma}_{d,\sigma}^{0}(\omega) = -\; \frac{1}{2} 
\left[ \begin{array}{cc}  
i \Gamma_{N} & \Gamma_{S} \\  
\Gamma_{S} & i \Gamma_{N}   
\end{array} \right]  
\label{static}
\end{eqnarray} 

Upon neglecting the Coulomb correlations one can analytically determine 
the Green's  function (\ref{GF}), where the spin dependent spectral 
function (\ref{rho_sigma}) acquires the BCS structure \cite{Tanaka-07}
\begin{eqnarray} 
\rho_{d,\sigma}(\omega) &=& \frac{1}{2} \left[ 1 + 
\frac{\varepsilon_{d}}{E_{d}} \right]  \frac{\frac{1}{\pi} \;  
\Gamma_{N}/2}{(\tilde{\omega}_{\sigma}\!-\!E_{d})^{2}+ 
(\Gamma_{N}/2)^{2}} \nonumber \\ &+& \frac{1}{2}  
\left[ 1 - \frac{\varepsilon_{d}}{E_{d}} \right]  
\frac{\frac{1}{\pi} \;\Gamma_{N}/2} 
{(\tilde{\omega}_{\sigma}\!+\!E_{d})^{2} 
+(\Gamma_{N}/2)^{2}} 
\label{rho_ingap} 
\end{eqnarray} 
with a quasiparticle energy $E_{d}=\sqrt{\varepsilon_{d}^{2}
+(\Gamma_{S}/2)^{2}}$. The in-gap QD states (often referred as 
{\em Andreev bound states}) form around $\pm E_{d} \! \pm \! 
\mu_{B}B$ as illustrated in the upper panel of figure 
\ref{bound_states}. Their line broadening is given by 
$\Gamma_{N}/2$ and in absence of magnetic field the particle-hole 
splitting is controlled by $\Gamma_{S}$  \cite{Tanaka-07,Domanski-07} 
(the dashed line in figure \ref{bound_states}). Magnetic field 
further enforces the Zeeman splitting of these in-gap states.

Above mentioned behavior has an indirect effect on the 
off-diagonal parts of the Green's function (\ref{GF}) which 
in turn determine the Andreev transmittance. In the 
limit $\Delta \rightarrow \infty$ (\ref{T_A}) reduces to  
\begin{eqnarray} 
T_{A,\sigma}(\omega) = \frac{\Gamma_{N}^{2}  
\left( \Gamma_{S}/2 \right)^{2} } 
{\left[ (\tilde{\omega}_{\sigma}\!-\!E_{d})^{2} 
\!+\!(\Gamma_{N}/2)^{2} \right] \left[ 
(\tilde{\omega}_{\sigma}\!+\!E_{d})^{2} 
\!+\!(\Gamma_{N}/2)^{2}\right] } 
\nonumber \\
\label{T_A_ingap} 
\end{eqnarray} 
The subgap Andreev conductance $G_{A}(V)=\frac{d}{dV} I_{A}(V)$ 
is thus characterized by a four peak structure  as shown in 
the bottom panel of figure \ref{bound_states}. Obviously 
the weights of particle and hole peaks of the spectral 
function (\ref{rho_ingap}) as well as their weights in  
the Andreev transmittance (\ref{T_A_ingap}) depend on the QD 
level $\varepsilon_{d}$.  Variation of the Andreev conductance 
with respect to ($V$, $\varepsilon_{d}$) is plotted in figure 
\ref{GA_vs_ed}.  We can notice that optimal conditions for 
the subgap current occur when the QD level is located near 
the energy gap center, otherwise the proximity effect is 
less efficient.  

\begin{figure}[t] 
\epsfxsize=10cm \centerline{\epsffile{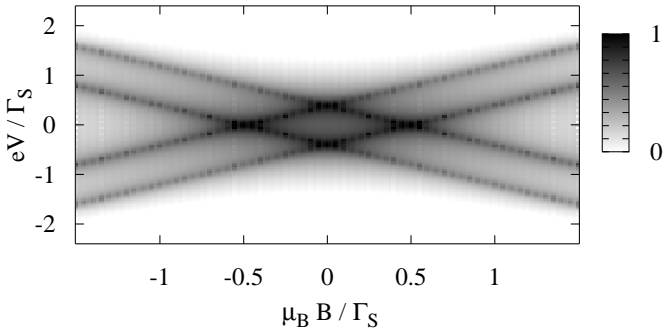}} 
\caption{Differential conductance $G_{A}(V)$ of the in-gap  
Andreev current as a function of the bias voltage $V$ and  
magnetic field $B$ for the QD level $\varepsilon_{d}\!=\!0$ and 
$\Gamma_{N}\!=\!0.1\Gamma_S$, $T=0.01\Gamma_{S}$ . 
Dark areas denote the regions where $G_{A}$ approaches  
the value $4e^{2}/h$.} 
\label{GA_vs_B} 
\end{figure} 
 
On top of the particle-hole structure seen in the Andreev states 
there is an additional Zeeman splitting brought by magnetic field. 
In figure \ref{GA_vs_B} we sketch the Andreev conductance in 
($V$,$B$) plane for $\varepsilon_{d}\!=\!0$, where the dark areas 
correspond to a maximal value $4e^{2}/h$. There appears a 
characteristic diamond shape marking the positions of such 
maximal conductance $G_{A}(V,B)$. We believe that this 
hyperfine structure could be probed experimentally. 
 
\begin{figure} 
\epsfxsize=8cm \centerline{\epsffile{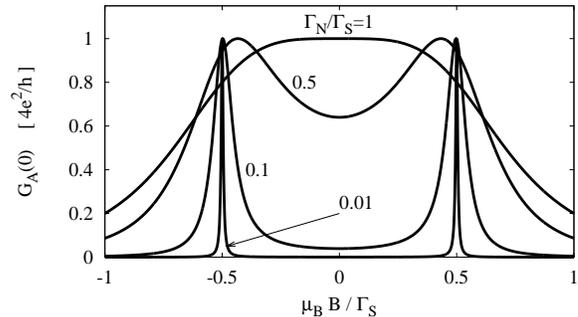}} 
\caption{The zero bias differential conductance $G_{A}(0)$  
as a function the magnetic field $B$ for several relative  
values of $\Gamma_{N}/\Gamma_{S}$. We used for computations 
$\varepsilon_{d}\!=\!0$ assuming $T\!\rightarrow\!0$.} 
\label{GA_zeroV} 
\end{figure}

To complete the discussion of the subgap Andreev current 
we briefly comment on a possible influence of an asymmetry 
between the hybridization couplings $\Gamma_{N}$, $\Gamma_{S}$. 
We explore for this purpose the zero bias conductance $G_{A}
(V\!=\!0)$. At low temperature we find from equation 
(\ref{T_A_ingap}) that   
\begin{eqnarray} 
&&G_{A}(0) = \label{GA_zero} \\&& \frac{4e^{2}}{h} \frac{\Gamma_{N}^{2} \;  
\left( \Gamma_{S}/2 \right)^{2} }{\left[ (\mu_{B}B 
\!-\!E_d)^{2}\!+\!\left( \frac{\Gamma_{N}}{2} \right)^{2}  
\right]\!\left[(\mu_{B}B\!+\!E_d)^{2}\!+\!\left(  
\frac{\Gamma_{N}}{2}\right)^{2}\right] } .
\nonumber  
\end{eqnarray} 
In figure \ref{GA_zeroV} we show the influence of magnetic 
field on the zero bias Andreev conductance for several 
values of the asymmetry rate $\Gamma_{N}/\Gamma_{S}$. If 
$\Gamma_{N}/\Gamma_{S}\! \ll \! 1$ then a line-broadening 
of the Andreev states diminishes so in consequence the 
particle and hole peaks become well separated. Under such  
conditions the subgap conductance has maxima around the 
quasiparticle states at $\pm\Gamma_{S}/2$ (where the ideal 
conductance $4e^{2}/h$ is reached). Let us recall, that in 
absence of magnetic field the equation (\ref{GA_zero}) 
reproduces for $\varepsilon_{d}\!=\!0$ the well known result 
$G_{A}(0)\! =\! \frac{4e^{2}}{h} \left( \frac{2\Gamma_{N} 
\Gamma_{S}}{\Gamma_{S}^{2}\!+\! \Gamma_{N}^{2}}\right)^{2}$ 
\cite{Tanaka-07}. For the symmetric coupling  $\Gamma_S
\!=\!\Gamma_N$ it yields $G_{A}(0)\! = \! 4e^{2}/h$ 
\cite{Fazio-98}. 

\section{Influence of the Coulomb correlations} 

In the limit $\Delta\!\rightarrow\!\infty$ the selfenergy 
${\mb \Sigma}_{d,\sigma}^{0}$ becomes a static quantity (\ref{static})
therefore the role of superconducting lead can be exactly replaced 
by the on-dot gap parameter $\Delta_{d}=\Gamma_{S}/2$. Instead of
(\ref{model}) we can thus use the following auxiliary Hamiltonian
\begin{eqnarray}
\hat{H} &=& \hat{H}_{N} + \sum_{{\bf k},\sigma } 
\left( V_{{\bf k} N} \; \hat{d}_{\sigma}^{\dagger}  
\hat{c}_{{\bf k} \sigma \beta } + \mbox{h.c.} \right) 
+ \sum_{\sigma} \epsilon_{d,\sigma} \hat{d}^{\dagger}_{\sigma} 
\hat{d}_{\sigma}  \nonumber \\ & + &
\left( \Delta_{d} \hat{d}_{\uparrow}^{\dagger}
\hat{d}_{\downarrow}^{\dagger} + \mbox{h.c.} \right)
+  U \; \hat{n}_{d \uparrow} \hat{n}_{d \downarrow} , 
\label{Tanaka_model}
\end{eqnarray}
which turns out to be very convenient for investigating the 
correlations. Tanaka and coworkers \cite{Oguri-04,Tanaka-07} were 
able to rigorously prove that the selfenergy ${\mb \Sigma}_{d,
\sigma}^{U}$ must have a diagonal structure due to invariance 
of $U \hat{n}_{d \uparrow} \hat{n}_{d \downarrow}$ term on 
the Bogoliubov-Valatin transformation.

In the remaining part of this section we shall focus on the subgap 
Andreev current transmitted through the correlated QD. The matrix 
Green's function (\ref{GF}) simplifies in the limit $\Delta \! 
\rightarrow \!\infty$  to the following (exact) structure  
\begin{eqnarray} 
&& {\mb G}_{\sigma}(\omega)= \label{arbitrary_U} \\
&&\left( \begin{array}{cc} \omega\!-\!\varepsilon_{d,\sigma}\!-\!
\Sigma_{N,\sigma}(\omega) &  \frac{1}{2}\Gamma_{S} \\ 
\frac{1}{2}\Gamma_{S} &  \omega\!+\!\varepsilon_{d,-\sigma}
\!+\!\Sigma_{N,-\sigma}^{*}(-\omega)\end{array}\right)^{-1} .
\nonumber
\end{eqnarray} 
Influence of the correlations have been so far analyzed for 
the Hamiltonian (\ref{model}) using various techniques 
\cite{Fazio-98,Clerk-00,Cuevas-01,Krawiec-04,Oguri-04,Tanaka-07,
Domanski-07}. Here we estimate the diagonal selfenergy $\Sigma_{N,
\sigma}(\omega)$ within (\ref{Tanaka_model}) by the equation 
of motion method \cite{EOM,TDAD-08}
\begin{widetext}
\begin{eqnarray}
\omega\!-\!\varepsilon_{d,\sigma}\!-\!\Sigma_{N,\sigma}(\omega)\!=\! 
\frac{[\omega\!-\!\varepsilon_{d,\sigma}\! -\!\Sigma^{0}_{d,\sigma}
(\omega)][\omega\!-\!\varepsilon_{d,\sigma}\! -\!U\!-\!
\Sigma^{0}_{d,\sigma}(\omega)\!-\!\Sigma^{3}_{d,\sigma}(\omega)]
\!+\!U \Sigma^{1}_{d,\sigma}(\omega)} {\omega-\varepsilon_{d,\sigma} 
- \Sigma^{0}_{d,\sigma}(\omega)-\Sigma^{3}_{d,\sigma}(\omega)
-U[1-\langle \hat{n}_{d,-\sigma}\rangle ]}
\label{ansatz}
\end{eqnarray}
where $\Sigma^{\nu=1,3}_{d,\sigma}(\omega)$  are 
given by \cite{EOM}
\begin{eqnarray}
\Sigma^{\nu}_{d,\sigma}(\omega) & = &  \sum_{\bf k} 
|V_{{\bf k} N}|^{2} \left(\frac{1}{\omega\!+\!\xi_{{\bf k} N}
\!-\!\varepsilon_{d,-\sigma}\!-\!\varepsilon_{d,\sigma}\!-\! U} 
+ \frac{1}{\omega\!-\!\xi_{{\bf k} N}\!+\!\varepsilon_{d,-\sigma}
\!-\!\varepsilon_{d,\sigma}} \right) \; \left[ f(\omega,T) 
\right]^{\frac{3 - \nu }{2}}  .
\label{sigma1and3}
\end{eqnarray}
\end{widetext}

\begin{figure} 
{\epsfxsize=9cm \centerline{\epsffile{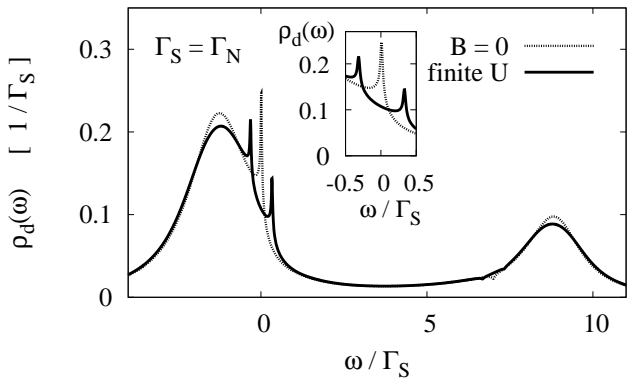}}} 
\caption{Spectral function of the correlated QD obtained for 
$\varepsilon_{d}=-1.5\Gamma_{S}$, $U\!=\!10\Gamma_{S}$,  
$\Gamma_{N} \! = \! \Gamma_{S}$ and temperature $T=10^{-3} 
\Gamma_{S}$ ($<< T_{K}$) in the limit $\Delta \! \rightarrow 
\! \infty$. Solid line corresponds to $\mu_{B}B\!=\!
\Gamma_{S}/3$.} 
\label{Kondo_peak} 
\end{figure}

Approximation (\ref{ansatz},\ref{sigma1and3}) qualitatively 
reproduces the following properties caused by on-dot correlations: 
(i) the charging effect and (ii) a possible appearance of the Kondo 
resonance for temperatures smaller than 
$T_{K}\!=\!\frac{\sqrt{U\Gamma_{N}}}{2} \mbox{exp}\{ \pi \varepsilon_{d}
\left( \varepsilon_{d} \! + \! U \right) / U \Gamma_{N}\}$. The 
latter one is related to screening of the quantum dot spin by 
itinerant electrons of the metallic lead. In the case when energy 
level $\varepsilon_{d}$ is located slightly below $\mu_{N}$ the 
hybridization $V_{{\bf k}N}$ induces effectively antiferromagnetic 
interaction between the QD and metallic lead. In consequence the 
bound singlet state can be formed giving rise to the resonance at 
$\omega=\mu_{N}$ for temperatures $T \leq T_{K}$. Magnetic field 
eventually splits this resonance as illustrated in figure 
\ref{Kondo_peak}.

Any features present in the QD spectrum are further showing 
up in the measurable differential conductance. This is also  
valid for the Kondo resonance. Since it forms near the chemical 
potential $\mu_{N}$ therefore its signatures appear predominantly 
in the low voltage current. In fact, it has been shown that Kondo 
resonance enhances at low temperatures the zero bias Andreev 
conductance \cite{Fazio-98,Domanski-07}, however its magnitude 
remains much smaller than the unitary limit value $2e^{2}/h$ 
typical for N-QD-N systems in the Kondo regime. In the present
context we emphasize that magnetic field enforces the Zeeman
splitting of the zero bias Andreev anomaly in much the same 
way as it affects the zero bias anomaly for the QD coupled 
to both metallic leads \cite{magn_N-QD-N,Kastner_etal}.

\begin{figure} 
{\epsfxsize=9cm \centerline{\epsffile{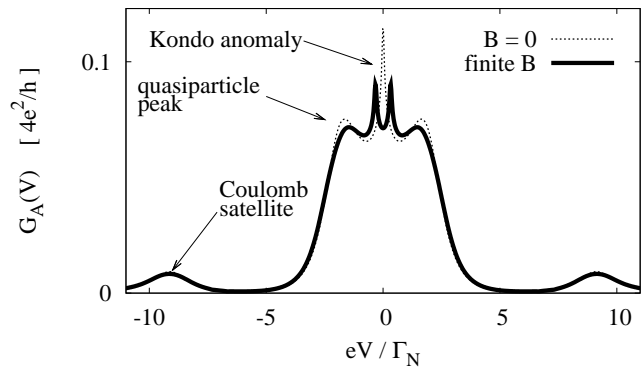}}} 
\caption{Effect of magnetic field on the subgap Andreev 
conductance $G_{A}(V)$ in the Kondo regime with $\varepsilon_{d}
\!=\!-1.5 \Gamma_{N}$, $U\!=\!10\Gamma_{N}$, $\Gamma_{S}\!=\!
5 \Gamma_{N}$ and $T \!=\! 10^{-3} \Gamma_{N} \ll T_{K}$.
Notice appearance of: (i) the zero-bias Kondo anomaly 
(showing the Zeeman splitting for $\mu_{B}B\!=\!\Gamma_{N}/3$, 
(ii) the quasiparticle peaks at $|eV|\! \simeq \!E_{d}$, and 
(iii) Coulomb satellite peaks near $|eV| \simeq U$.} 
\label{anomaly} 
\end{figure} 

\begin{figure} 
{\epsfxsize=9.5cm \centerline{\epsffile{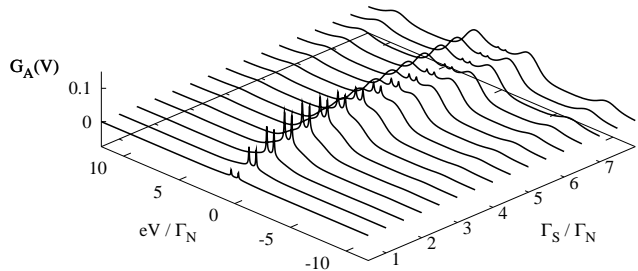}}} 
\caption{The differential Andreev conductance $G_{A}(V)$ 
(in units $4e^{2}/h$) as a function of the bias voltage $V$ 
and the asymmetry ratio $\Gamma_{S}/\Gamma_{N}$. We used 
the same set of parameters as in figure \ref{anomaly}.} 
\label{opt_conditions} 
\end{figure}

The zero bias enhancement of the Andreev conductance is a feature 
whose presence might be difficult to notice \cite{Fazio-98,Clerk-00,
Domanski-07} unless some stringent requirements are fulfilled 
\cite{TDAD-08}. It turns out that optimal conditions for the low 
temperature enhancement of $G_{A}(V\!\sim\!0)$ take place when 
$\Gamma_{S}$ is comparable to $\Gamma_{N}$ (see figure 
\ref{opt_conditions}) and $\varepsilon_{d}$ is located slightly 
below the energy gap center. For an increasing asymmetry between 
the hybridizations $\Gamma_{N}$, $\Gamma_{S}$ 
the magnitude of low voltage Andreev conductance diminishes 
(similarly as we have been shown in section IV upon neglecting 
the correlations). On the other hand, for $\varepsilon_{d}$ moving 
far aside from the superconductor's gap center the proximity 
effect becomes weakened and the overall Andreev conductance 
is again suppressed. 

In general it seems that an interplay between the on-dot pairing 
(absorbed from the superconducting electrode) and the Kondo state 
(due to screening of QD spin by the metallic lead electrons) has 
the same character as a competition of superconductivity versus 
magnetism in the solid state physics.  Since this is outside 
the main scope of the present topic we shall discuss it 
separately \cite{TDAD-08}. A combination of the Kondo physics, 
superconductivity and the Zeeman polarization is a complex 
problem and to our knowledge only few papers have so far 
attempted to address this challenging issue \cite{Yamada-07}.   

\section{Summary}
 
We have explored the effect of magnetic field on charge 
transport through the quantum dot attached to one normal 
and one superconducting electrode. For a bias voltage 
$V\!\simeq\!\pm\Delta/e$ we find the Zeeman splitting  
of the square root singularities in the differential  
conductance. This resembles the experimental result  
of Meservey, Tedrow and Fulde observed in the N-I-S  
junction  \cite{Meservey-70} which for the N-QD-S 
structures it seems rather easy to achieve. 
 
We have extended our study also on the in-gap Andreev current.
Due to the proximity effect the particles and holes of the 
quantum dot get mixed and effectively the spectrum acquires 
the BCS-like structure (\ref{rho_ingap}). Differential 
conductance $G_{A}(V)$ of the in-gap current indirectly probes 
such structure of the bound Andreev states. We have shown 
that magnetic field leads to appearance of four peaks via 
the combined particle-hole and Zeeman splittings. We hope 
that this result might stimulate a search for the experimental 
detection of above mentioned structures.
 
Moreover, we have explored influence of the on-dot Coulomb 
interactions on the subgap Andreev current assuming the 
extreme limit $\Delta\!\rightarrow\!\infty$. In general, 
the on-dot correlations contribute to the QD spectrum: 
(i) appearance of the Coulomb satellite near $\omega\!=\!
\varepsilon_{d,\uparrow}+\varepsilon_{d,\downarrow}+U$ 
(charging effect), and (ii) at sufficiently low temperatures 
can produce the narrow Kondo resonance at the chemical potential 
$\mu_{N}$. Magnetic field imposes the hyperfine splitting onto 
such spectrum in a similar way as has been observed in N-QD-N junctions 
\cite{Kastner_etal}. The Kondo effect alone is exemplified 
in the zero bias Andreev conductance where under appropriate 
conditions \cite{TDAD-08} a low temperature enhancement can 
be seen if $\Gamma_{N} \sim \Gamma_{S}$ and the gate voltage 
tunes $\varepsilon_{d}$ nearly to the energy gap center.
 
It would be of interest to use some more sophisticated methods 
for treating the on-dot interaction $U$ in order to check 
whether there exist a minimal magnetic field necessary for 
splitting the Kondo peak (as theoretically predicted for N-QD-N 
junctions \cite{magn_N-QD-N}) observable in the Andreev conductance.
One can also study QD coupled with $d$-wave superconductor, 
where the square root singularities are replaced by weaker 
kinks. We think that the Meservey-Tedrow-Fulde effect would 
be observable there too (but in a less pronounced manner) 
whereas the subgap conductance might qualitatively change.

\vspace{0.5cm} 
{\em Acknowledgment} 
We thank A.M.\ Gabovich for discussion leading to this 
study and P. Fulde for interest and encouragement.
This work  was partly supported by the Ministry of Science 
and  Education under the grants NN202187833 and NN202373333.

\end{document}